\documentclass[12pt]{iopart}
\usepackage{iopams}  
\usepackage{cite}
\usepackage[english]{babel}

\newcounter{theo}

\newcounter{prop}

\newcounter{rem}
\newcommand{\rem}{\refstepcounter{rem}\noindent\textbf{Remark \therem.} }
\newcounter{exam}

\newcounter{lem}

\newcounter{defi}

\begin{document}

\title[Integrable non-abelization of the flow on an elliptic curve]{Integrable non-abelization of the flow on an elliptic curve}

\author{V.\ Sokolov $^{abc}$ and T.\ Wolf $^{d}$}

\address{$^a$ Landau Institute for Theoretical Physics, Chernogolovka, Russia\\
$^b$ Centre of Integrable Systems, P.G. Demidov Yaroslavl State
University, Russia \\
$^c$ UFABC, Sao Paulo, Brasil \\
$^d$ Department of Math \& Stat, Brock University, St Catharines, ON, Canada}
\ead{vsokolov@landau.ac.ru, twolf@brocku.ca}
\begin{abstract}
An integrable non-abelian generalization of a Hamiltonian flow on an elliptic curve is presented. A Lax pair for this non-abelian system is found.
\end{abstract}

\pacs{02.30.Ik (Integrable systems), 02.30.Hq (Ordinary differential equations)}
\vspace{2pc}
\noindent{\it Keywords}: Integrability, non-abelian ODE systems, flows on elliptic curves.\ \ \ 

\bigskip

Let us consider the following system of ODEs: 
\begin{equation}
\left\{
\begin{array}{lcl}
u_t   =  v^2 + c u + a \, {\rm I},\\[2mm]
v_t   =  u^2 - c v + b \, {\rm I}\label{inuv}
\end{array} ,  
\right.
\end{equation}where $u$ and $v$ are $m\times m$-matrices, ${\rm I}$ is the identity matrix, $a , b$ and $c$ are arbitrary constants. It was observed in \cite{sokwolf} that this non-abelian system has many polynomial integrals and infinitesimal symmetries for any $m$.   It is easy to see that all components of the matrix $uv - vu$ are first integrals for the non-abelian system.

In the case $m=1$ we have a system of two ODEs which can be written in the Hamiltonian form 
$$
u_t = - \frac{\partial H}{\partial v}, \qquad v_t =  \frac{\partial H}{\partial u}
$$
with the Hamiltonian 
$$
H = \frac{1}{3} u^3 - \frac{1}{3} v^3 - c u v + b u - a v.
$$
For generic $a,b,c$ the relation $H = const$ is an elliptic curve   and (\ref{inuv}) describes the motion of  its point. 

In the case of arbitrary $m$ the system (\ref{inuv}) remains to be Hamiltonian with the Hamiltonian
\begin{equation}\label{HAMmat}
H = {\rm tr}\, \left( \frac{1}{3} u^3 - \frac{1}{3} v^3 - c u v + b u - a v \right)
\end{equation} 
and non-abelian constant Poisson bracket \cite{odrubsok}.

System (\ref{inuv}) admits the following obvious discrete transformations, which change the values of the parameters $a, b, c$:
\begin{eqnarray*}
\hspace*{-20pt}R_1 &:&  \qquad \bar u = \varepsilon u,\qquad \bar v = \varepsilon^2 v, \qquad \bar a = \varepsilon a, \qquad \bar b = \varepsilon^2 b, \qquad \bar c = c ;\\ [3pt]
\hspace*{-20pt}R_2 &:&  \qquad \bar u = v,\qquad \bar v = u, \qquad \bar a = b, \qquad \bar b = a, \qquad \bar c = -c.  
\end{eqnarray*} 
In the scalar case V. Adler \cite{adler} found one more transformation
$$\hspace*{-20pt}R_3 :  \qquad \bar u = u + \frac{a-b}{v-u+c},\qquad \bar v =  v + \frac{a-b}{v-u+c}, \qquad \bar a = b, \qquad \bar b = a, \qquad \bar c = c. 
$$
It is easily verified that $R_1^3=R_2^2=R_3^2 = {\bf Id}.$ The transformation $R_1 R_3 R_1 R_3$ does not change the parameters and  defines a non-trivial shift on the elliptic curve $H = {\rm const}.$ It can be verified that the transformation $R_3$ is applicable in the matrix case as well.

\section{Lax representation}

In the homogeneous non-abelian case $c=a=b=0$ (see \cite{miksokcmp}) the following Lax $(L,A)$-pair 
\begin{equation}\label{HomLax}\begin{array}{c}
L = \left(\begin{array}{ccc} 1 & 0 & 0\\ 0 & \varepsilon & 0\\0 & 0 & \varepsilon^2 \end{array}\right) \, \lambda + 
\left(\begin{array}{ccc} 0 & 3 \varepsilon u & 3 v \\ v & 0 & ( \varepsilon -1) u\\u& (2 \varepsilon+1) v & 0 \end{array}\right), \\ \\
\displaystyle A = - \frac{1}{3} \left(\begin{array}{ccc} \varepsilon^2 & 0 & 0\\ 0 & \varepsilon & 0\\ 0 & 0 & 1 \end{array}\right) \, \lambda + 
\frac{1}{3} \left(\begin{array}{ccc} 0 & 3 \varepsilon^2 u & 3 v\\ \varepsilon v & 0 & ( \varepsilon +2) u\\u& (1-\varepsilon) v&0 \end{array}\right)
\end{array}
\end{equation}
where 
\begin{equation}\label{eps}
\varepsilon^2+\varepsilon+1=0,
\end{equation}
can be derived from Section 3.1 of \cite{miksokcmp} and from the Lax pair 
$$
L = \lambda\,C + M, \qquad A = \frac{1}{\lambda} \, M^2
$$
of the non-abelian Manakov's equation  \cite{man}
$$M_t = [M^2, \, C].$$ 
 In this equation $M(t)$ is an unknown matrix of arbitrary size and $C$ is a given constant matrix. 

{\bf Proposition 1.} The Lax equation 
\begin{equation} \label{Lax}
\bar L_t = [A,\,\bar L],
\end{equation}
where
\begin{equation}\label{InHomLax} 
\bar L = \lambda\, L + \lambda\, c\, P + a\,Q + b\, R,
\end{equation}
\medskip
{\small
$$
P=\left(\begin{array}{ccc}\varepsilon+2&0&0\\0&-2\varepsilon -1&0\\0&0&\varepsilon-1\end{array}\right),\, 
Q=\left(\begin{array}{ccc}0&3(\varepsilon+2)&0\\0&0&-3\\ \varepsilon -1&0&0\end{array}\right),\,
R=\left(\begin{array}{ccc}0&0&3(1-\varepsilon)\\2 \varepsilon+1&0&0\\0&-3\varepsilon&0\end{array}\right)
$$
and $L, A$ are given by (\ref{HomLax}), is equivalent to the non-abelian system (\ref{inuv}).

{\bf Remark.} It is easy to see that formulas (\ref{Lax}) and (\ref{InHomLax}) define also the Lax representation for (\ref{inuv}) with $u$ and $v$ being elements of any associative algebra. 

\section{Trace first integrals}

\qquad  Consider  first integrals of system  (\ref{inuv}) of the form ${\rm tr}\,(P(u,v))$, where $P$ is a polynomial. We called $P$ a {\it trace first integral}. For the polynomial $P$ we use the abbreviation TFI. The TFI is defined up to any linear combination of commutators of matrix polynomials.

As it was mentioned, system (\ref{inuv}) has the matrix first integral $vu-uv$ and therefore the trace of each one of its powers $(vu-uv)^n$ is a trace first integral $M_n$ of degree $2n$. For example,
\[ M_2 = v^2u^2 - vuvu \]
\[ M_3 = v^2u^2vu - v^2uvu^2 \]
\[ M_4 = v^2u^2v^2u^2 - 2v^2u^2vuvu + 2v^2uvu^2vu - 2v^2uvuvu^2 + vuvuvuvu .\]

All other TFI come from the Lax representation.
Namely, it follows from (\ref{Lax}) that
\begin{equation}\label{trr}
\Big({\rm tr}\, (\bar L^k)\Big)_t = 0
\end{equation}
for any $k.$  Each expression ${\rm tr}\, (\bar L^k)$ is a polynomial in $\lambda$, whose 
coefficients are first integrals. All these TFI involve parameters $a, b, c.$ 

Moreover, replacing $\varepsilon^2$ by $-\varepsilon-1$ 
the resulting expressions are linear in $\varepsilon.$ Since $\varepsilon$ is any of two solutions of the quadratic equation (\ref{eps}), we obtain two TFI through the 
coefficients of $\varepsilon^1, \varepsilon^0$. 

For $k=1,2$ these integrals are trivial. In the case $k=3$ the Hamiltonian (\ref{HAMmat}) 
arises. Relations (\ref{trr}) with $k=4,\dots,9$ produce the integrals ${\rm tr}\, (T_i)$ , where

\begin{eqnarray*}
\hspace*{-20pt}T_1&\hspace*{-10pt}=&v^6 - 6v^3u^3 + 6v^2uvu^2 - 2vuvuvu + u^6 + 6v^4uc - 6vu^4c + 6v^4a - 6vu^3a - 6v^3ub\\
 & &+ 6u^4b + 9vuvuc^2 + 18v^2uac - 18vu^2bc - 18vuab + 9v^2a^2 + 9u^2b^2,\\[2mm]
\hspace*{-20pt}T_2&\hspace*{-10pt}=&v^5u^2 - 2v^4uvu + v^3uv^2u + 2vu^4vu - v^2u^5 - vu^3vu^2 + 3v^2uvu^2c - 3vuvuvuc\\
 & &+ 3v^3u^2a - 3v^2uvua - 3v^2u^3b + 3vu^2vub,\\[2mm]
\hspace*{-20pt}T_3&\hspace*{-10pt}=&v^5u^2vu - v^5uvu^2 - v^4u^2v^2u + v^4uv^2u^2 - v^3uv^2uvu + v^3uvuv^2u - v^2u^5vu\\
 & &+ v^2u^4vu^2 - v^2u^2vu^4 + v^2uvu^5 - vu^3vu^2vu + vu^3vuvu^2 + 3v^2u^2vuvuc - 3v^2uvuvu^2c\\
 & &+ 3v^3u^2vua - 3v^3uvu^2a + 3v^2uvu^3b - 3v^2u^3vub,\\[2mm]
\hspace*{-20pt}T_4&\hspace*{-10pt}=&v^9 - 9v^6u^3 + 9v^5uvu^2 + 9v^4u^2v^2u - 9v^4uvuvu + 9v^3u^6 - 9v^3u^2v^3u + 9v^3uv^2uvu\\
 & &- 9v^2u^4vu^2 + 9v^2u^3vu^3 - 3v^2uv^2uv^2u - 9v^2uvu^5 + 9vu^4vuvu + 3vu^2vu^2vu^2\\
 & &- 9v^3u^3vuc - 9v^3u^2vu^2c - 9v^3uvu^3c - 27v^3uvubc - 9v^2u^3v^2uc + 9v^2u^2vuvuc\\
 & &+ 36v^2uvuvu^2c + 9vu^7c + 81vu^3b^2c - 18vuvuvuvuc - 9v^3u^2vua + 18v^3uvu^2a\\
 & &+ 18v^2u^2v^2ua - 9v^2uvuvua + 9vu^6a + 36v^3u^4b + 9v^2u^3vub - 18v^2u^2vu^2b\\
 & &- 18v^2uvu^3b + 9vu^2vuvub - 9u^7b - 27v^2u^4ac - 27v^2u^3a^2 + 27v^3u^2b^2\\
 & &+ 54vu^5bc - 27vu^4vuc^2 + 54vu^4ab - 27vu^3vuac - 27u^5b^2 + 81v^3ua^2c\\
 & &+ 81v^2uvuac^2 - 81v^2ua^2b - 81vu^2vubc^2 + 81vu^2ab^2 + 27vuvuvuc^3 - 162vuvuabc\\
 & &+ 27u^3a^3 - 27u^3b^3 - 81vua^3c - 81va^4 + 81ua^3b.
\end{eqnarray*}

We verified that any TFI of degree not greater that 9  is a linear combination of $H, T_1,...T_4$ and $M_2,..., M_4$ up to commutators.

According to Yu.\ Suris \cite{yusu} these trace integrals together with the components of the matrix $u v - v u$ provide a complete set of functionally independent integrals in the cases of $2\times 2$ and $3\times 3$ - matrices.

\rem The TFIs resulting from splitting ${\rm tr}\, (\bar L^k)$ are not necessarily in their shortest form.  We reduced the number of terms in TFI by subtracting of lower degree and same degree TFIs as well as commutators of monomials. To do that we implemented a special algorithm in the computer algebra system REDUCE.

The following table shows how many new trace first integrals of degree $d$ appear in the coefficient $\lambda^m$ in the power tr$\, (\bar L^k)$ as the exponent $k$ is increased. Brackets () indicate traces of the matrix first integrals which are homogeneous and free of $a,b,c$. Brackets [] indicate inhomogeneous first integrals which involve $a,b,c$.
\begin{center}
 \begin{tabular}
  {| c |  c  |c|c|  c  |c|  c  |  c  |c|  c  |  c  | c|   c   |   c   |}   \hline
   k  :&  0  &1&2&  3  &4&  5  &  6  &7&  8  &  9  &10&   11  &   12    \\ \hline
 (m,d):&(0,0)& & &[3,3]& &(6,4)&(6,6)& &[9,7]&(9,8)&  &(12,10)&(12,12)  \\ \hline 
       &     & & &     & &     &[6,6]& &     &[9,9]&  &[12,10]&[12,12]  \\ \hline 
       &     & & &     & &     &     & &     &[9,9]&  &       &         \\ \hline 
 \end{tabular}
 $ $\vspace{6pt}
 \begin{tabular}
  {| c |   c   |   c   |  c    |   c   |   c   |   c   |   c   |   c   |}   \hline
   k  :&   13  &   14  &  15   &   16  &   17  &   18  &   19  &   20    \\ \hline
 (m,d):&[15,11]&[15,13]&(15,14)&[18,14]&(18,16)&[18,18]&[21,17]&[21,19]  \\ \hline
       &       &[15,13]&[15,15]&       &[15,16]&[18,18]&[21,17]&[21,19]  \\ \hline
       &       &       &[15,15]&       &[15,16]&[21,15]&       &         \\ \hline
 \end{tabular}
$ $ \\ $ $ \\ Table 1. Occurrences of new TFI of degree $d$ in the coefficient of $\lambda^m$ in  tr$\, (\bar L^k)$
\end{center}

We believe that the integrable non-abelian system (\ref{inuv}) could be one of the keys on which a theory of non-abelian elliptic functions can be built.
\medskip
 
 {\bf Acknowledgments.} The authors are grateful to V. Adler,  M. Kontsevich and Yu. Suris for useful discussions.
The first author is grateful to IHES for hospitality and support. He was partially supported by RFBR grant 16-01-00289.

\end{document}